\documentstyle[twoside,fleqn,espcrc2]{article}

   \newcommand{\cM}{{\cal M}}
   \newcommand{\cA}{{\cal A}}
   \newcommand{\cL}{{\cal L}}
   \newcommand{\cG}{{\cal G}}
   \newcommand{\cC}{{\cal C}}

\newcommand{\AmS}{{\protect\the\textfont2
  A\kern-.1667em\lower.5ex\hbox{M}\kern-.125emS}}

\newcommand{\IZ}
{{\bf Z}}
\newcommand{\IR}
{{\bf R}}

\def\balpha{ {\mbox{\boldmath $\alpha$}}}
\def\bbeta{ {\mbox{\boldmath $\beta$}}}
\def\bgamma{ {\mbox{\boldmath $\gamma$}}}
\def\hK{hyper-K{\" a}hler}

\newcommand\bea{\begin{eqnarray}}
\newcommand\eea{\end{eqnarray}}
\def\nonu{\nonumber}

\def\Tr{{\rm Tr}}
\def\a{\alpha}
\def\b{\beta}
\newcommand{\beq}{\begin{equation}}
\newcommand{\eeq}{\end{equation}}

\newcommand{\n}{\nu}
\newcommand{\m}{\mu}

\newcommand{\eps}{\epsilon}
\newcommand{\vP}{{\bf P}}
\newcommand{\vZ}{{\bf Z}}
\newcommand{\vn}{{\bf n}}
\newcommand{\vg}{{\bf g}}
\newcommand{\vh}{{\bf h}}
\newcommand{\vq}{\bf q}
\newcommand{\vH}{\bf H}
\newcommand{\p}{\partial}

\def\NP{{\it Nucl. Phys.\ }}
\def\AP{{\it Ann. Phys.\ }}
\def\PL{{\it Phys. Lett.\ }}
\def\PR{{\it Phys. Rev.\ }}
\def\PRL{{\it Phys. Rev. Lett.\ }}
\def\CMP{{\it Comm. Math. Phys.\ }}

\title{Duality and Supersymmetric Monopoles}

\author{Jerome P. Gauntlett \address{Physics Department, Queen Mary and 
Westfield College, \\Mile End Rd, London E1 4NS, U.K.}%
        \thanks{Current Address: Isaac Newton Institute, 20 Clarkson Rd,
                Cambridge, CB3 0EH.}
      } 
\begin{document}
\begin{abstract}
Exact duality in supersymmetric gauge theories leads to highly non-trivial
predictions about the 
moduli spaces of BPS monopole solutions. These notes 
attempt to be a pedagogical review of the current status
of these investigations and are based on lectures given at the 
33rd Karpacz Winter School: Duality - Strings and Fields, February 1997.
 
\end{abstract}

\maketitle

\section{INTRODUCTION}

Electromagnetic duality has emerged as a powerful tool to study strongly
coupled quantum fields. In $N$=4
super-Yang-Mills theory and some special theories with $N$=2
supersymmetry, the duality is conjectured to be exact
in the sense that it is valid at all energy scales. 
These theories provide the most natural setting for Montonen and
Olive's original idea \cite{mo} since they
have vanishing $\beta$-functions and hence the quantum
corrections are under precise control.
In many theories with $N$=2 and $N$=1 supersymmetry duality plays an important
role in elucidating the infrared dynamics. In these
models one can study strong coupling phenomenon such as
confinement and chiral symmetry breaking in an exact
context\footnote{See the contributions of other speakers at the school.}. 

The purpose of these lectures is to review some aspects of
exact duality focusing on theories with $N$=4 supersymmetry.
In these theories the duality group is $SL(2,\IZ)$ which
includes a $\IZ_2$ corresponding to the interchange of electric and magnetic
charges along with the interchange of strong and weak coupling.
Sen was the first to realise \cite{sen} that $SL(2,\IZ)$ or 
``$S$-duality" leads to highly non-trivial predictions
about the BPS spectrum of magnetic monopoles and dyons in the theory.
BPS states are important for testing duality because they
form short representations
of the supersymmetry algebra and hence we have good control over
their behaviour as we vary the coupling. 
At weak coupling the predicted BPS spectrum 
can be translated 
into statements about  
certain geometric structures on the moduli space of BPS monopole
solutions. 


We begin with bosonic $SU(2)$ BPS monopoles, reviewing 
some aspects of the moduli space approximation and discussing how
quantised dyons appear in the semiclassical spectrum.
Next we describe some features of $N$=4 super-Yang-Mills
theory before studying the $S$-duality predictions.
We analyse the
$SU(2)$ case followed by the higher rank gauge groups.
We conclude by outlining some open problems
in the study of exact duality.

\section{SU(2) BPS MONOPOLES AND THE MODULI SPACE APPROXIMATION}

Consider the Yang-Mills-Higgs Lagrangian
\beq
\cL= -{1\over 4}F^a_{\m\n}F^{a\m\n}-{1\over 2}
D_\m\Phi^a D^{\m}\Phi^a~,
\label{eq:one}
\eeq
where $A_\m=A^a_\mu T^a$ is an SU(2) connection with field strength
$F_{\m\n}=\p_\m A_\n-\p_\n A_\m+e[A_\m, A_\n]$, and $\Phi=\Phi^a T^a$ 
transforms in the adjoint representation
with the covariant derivative given by 
$D_\m\Phi=\p_\m\Phi+e[A_\m,\Phi]$. We choose the Lie algebra generators
$T^a$ to be anti-hermitian. 
There is no potential term for
the Higgs field and we are thus considering 
the ``BPS limit" \cite{ps,bog} 
which is relevant for the supersymmetric extension.
The moduli space of Higgs vacuua is obtained by imposing  
$\langle\Phi^a\Phi^a\rangle=v^2$ and is thus a two-sphere. If $v^2\ne0$
then $SU(2)$ is 
spontaneously broken 
to $U(1)$.
The electric and magnetic charge with respect to the $U(1)$
specified by the Higgs field are given by
\bea 
Q_e &=& {1\over v}\int d{S^i} ({E}_i^a \Phi^a) ~, \nonu \\
Q_m &=& {1\over v}\int d{S^i} ({B}_i^a \Phi^a) ~,
\label{eq:two}
\eea
where $E^a_i=F_{0i}$ and $B^a_i={1\over 2}
\eps_{ijk}F^a_{jk}$ are the non-abelian electric and
magnetic field strengths, respectively, and the integration is over a 
surface at spatial infinity.

The perturbative states consist of a massless photon, a massless
neutral scalar and
$W^\pm$ bosons with electric charge $Q_e=\pm e$ and mass $ev$.
To analyse the monopole and dyon spectrum we need to construct
classical static monopole solutions and then perform a semi-classical
analysis. 
Let us begin by noting that for all
finite energy configurations the Higgs field must lie in the vacuum
at spatial infinity. The Higgs field of these configurations 
thus provides a map from the two
sphere at spatial infinity to the two sphere of Higgs vacuua. These
maps are characterised by a topological winding 
number $k$ and one can show that this implies that the magnetic
charge is quantised
\beq
Q_m={4\pi\over e}k~.
\label{eq:three}
\eeq
The minimal magnetic monopole charge is twice the Dirac unit
because we could add electrically charged fields in the fundamental
representation of $SU(2)$ that would carry 1/2 integer electric charges
in contrast to the integer charged $W$-bosons.

To proceed with the construction of static monopole solutions
it
will be convenient to work in the $A_0=0$ gauge.
We must then impose
Gauss' Law, the $A_0$ equation of motion, as a constraint on
the physical fields:
%
\beq
D_i\dot A_i + e[\Phi,\dot\Phi]=0.
\label{eq:four}
\eeq
In this gauge the Hamiltonian is $H=T+V$ where the kinetic
and potential energies are given by
\bea
T&=&{1\over 2}\int d^3x(\dot A^a_i \dot A^a_i+\dot\Phi^a \dot\Phi^a)~,
\label{eq:five}
\\
V&=&{1\over 2}\int d^3x(B^a_iB^a_i+D_i\Phi^a D_i\Phi^a),
\eea
respectively.
Noting that $V$ can be rewritten \cite{bog} as
%
\beq
V={1\over 2}\int d^3x[(B^a_i\mp
D_i\Phi^a)(B_i^a\mp D_i\Phi^a)]\pm vQ_m,
\eeq
we deduce that in each topological class $k$ 
corresponding to magnetic charge given by (\ref{eq:three})
there is a  Bogomol'nyi bound on the mass of any static classical
monopole solution:
%
\beq 
M\ge v|Q_m|={4\pi v\over e}|k|.
\label{eq:eight}
\eeq
The static energy is minimised when the
bound is saturated which is equivalent to
the Bogomol'nyi (or BPS) equations
%
\beq
B_i=\pm D_i\Phi.
\eeq
The upper sign corresponds to positive $k$ or `` monopoles" and
the lower sign corresponds to negative $k$ or ``anti-monopoles". From
now on we will restrict our considerations to monopoles, the extension
to anti-monopoles being trivial. 
In the $A_0=0$ gauge there are no static dyon solutions; the dyons emerge
as time dependent solutions as we will see.

The moduli space of gauge inequivalent solutions to the Bogomol'nyi equations
will be denoted $\cM_k$. 
Let us discuss some of the geometry of this manifold.
We begin by recalling that in the $A_0=0$ gauge the
configuration space of fields is given by 
$\cC=\cA/\cG$ where $\cA=
\{A_i(x),\Phi(x)\}$ is the space of finite energy field configurations
and we have divided out by $\cG$, the group of gauge transformations
that go to the identity at spatial infinity. 
Tangent vectors $\{\dot A,\dot \Phi\}$ to $\cC$ must satisfy Gauss
Law (\ref{eq:four}). From 
this point of view, the kinetic energy in (\ref{eq:five}) 
is simply the metric on $\cC$. 
The moduli $Z^\alpha$ that appear in the general solution to the 
Bogomol'nyi equations
$\{A(x,Z),\Phi(x,Z)\}$, are natural coordinates on $\cM_k\subset \cC$. 
Tangent vectors to $\cM_k$ must
also satisfy the linearised Bogomol'nyi equations
\beq
\epsilon_{ijk}D_j\dot A_k=D_i\dot\Phi+e[\dot A_i,\Phi]~.
\eeq
Using the coordinates $Z^\alpha$ we have
\beq
\{\dot A, \dot \Phi\}=\dot Z^\alpha\{\delta_\alpha A_i,\delta_\alpha\Phi\}~,
\eeq
where $\{\delta_\alpha A_i,\delta_\alpha\Phi\}$ satisfy
\bea
&&D_i\delta_\alpha A_i +e[\Phi,\delta_\alpha\Phi]=0~,
\\
&&\eps_{ijk}D_j \delta_\alpha A_k=
D_i\delta_\alpha\Phi+e[\delta_\alpha A_i,\Phi]~,
\label{eq:thirteen}
\eea
which are simply the equations for a physical zero mode. 
The zero modes can be obtained by differentiating the general solution
with respect to the moduli but in general one has to include a gauge
transformation to ensure that it satisfies (\ref{eq:thirteen}) i.e.,
\bea
\delta_\alpha A_i&=& \p_\a A_i - D_i\eps_\alpha~,\nonu\\
\delta_\alpha\Phi&=&\p_\alpha\Phi-e[\Phi,\eps_\a]~.
\eea	
The metric on $\cC$
gives rise to a metric on $\cM_k$ which can be written in terms of
the zero modes:
\beq
\cG_{\alpha\beta}(Z)=\int d^3x[\delta_\alpha A^a_i\delta_\beta A^a_i
+\delta_\alpha\Phi^a\delta_\beta\Phi^a]~.
\label{eq:fifteen}
\eeq
$\cM_k$ is $4k$-dimensional which can be established, for example, 
by counting zero modes using an index theorem.
The space of field configurations $\cA$ inherits three almost
complex structures,
from those on $\IR^4$ and they descend to give a \hK\ 
structure on $\cM_k$. Explicit formulae for the complex structures on $\cM_k$
in terms of the zero modes can be found in \cite{gauntlett}. More
details on the geometry of $\cM_k$ can be found in \cite{atiyah}.

The moduli space for a single BPS monopole
can be determined by explicitly constructing the 
most general solution and we find 
$\cM_1=\IR^3\times S^1$. The $\IR^3$ piece simply
corresponds to the position of the monopole in space. The $S^1$ 
arises from the gauge transformation
$g=e^{\chi\Phi/v}$ on any solution. Since this does not go to
the identity at infinity, it is a ``large" gauge transformation, it
corresponds to a physical motion. Since all fields are in the adjoint
a $2\pi$ rotation in $SU(2)$ is the identity and 
we conclude that $0\le \chi<2\pi$. 
We will see that this coordinate is a dyon degree of freedom.

We have noted that the dimension of $\cM_k$ is $4k$. The physical
reason for the existence of these multimonopole
configurations is that in the BPS limit there is a cancellation
between the vector repulsion and scalar attraction between two
monopoles. Heuristically one can think of the $4k$
dimensions as corresponding to a position in $\IR^3$ and a phase for
each monopole but the structure of $\cM_k$ turns out to be much more
subtle and interesting. 
For general $k$ we can separate out a piece
corresponding to the motion of the centre of mass of the multi-monopole
configuration and we have $\cM_k=\IR^3\times 
(S^1\times \tilde {\cM}_k^0)/\IZ_k$.
The $S^1$ factor is related to the total electric charge.
$\tilde \cM^0_k$ is $4(k-1)$ dimensional and \hK. It admits 
an $SO(3)$ group
of isometries which corresponds to a rotation of the multi
monopole configuration in space. Although 
the topology of these spaces is well understood,
the metric is explicitly known only for $k=2$ \cite{atiyah}.

To determine the semi-classical spectrum of states with magnetic charge
$k$ we start with a classical solution $(A^{cl}(x,Z),\Phi^{cl}(x,Z))$.
To have a well defined
perturbation scheme with $e\ll1$, we need to introduce a collective co-ordinate
for each zero mode; these are the moduli
$Z^\a$. We then expand an arbitrary time dependent field
as a sum of the massive modes with time 
dependent coefficients and allow the collective coordinates to
become time-dependent 
(see, e.g., \cite{raj}).
A low-energy ansatz for the fields is obtained
by ignoring
the massive modes and demanding that the only time dependence is via the
collective co-ordinates. Thus we are led to the ansatz
\footnote{Note that the $A_0$ term is included to ensure that the motion
is orthogonal to gauge transformations.
One could do a gauge transformation if one 
wants to remain in the $A_0=0$ gauge (see also the discussion in \cite{wein}).}
%
\bea
A_i(x,t)&=& A^{cl}_i(x, Z(t))~,\nonu\\
\Phi(x,t)&=& \Phi^{cl}(x, Z(t)),\nonu\\
A_0&=&\dot Z^\alpha\eps_\alpha~.
\eea
After substituting this into the action (\ref{eq:one})
we obtain an effective action
\beq
S={1\over 2}\int dt\cG_{\a\b}\dot Z^\a \dot Z^\b-{4\pi v\over e} k~,
\label{eq:seventeen}
\eeq
which is precisely that of a free particle propagating on the
moduli space $\cM_k$ with metric (\ref{eq:fifteen}). This is the moduli
space approximation
\cite{man}. The classical equations of motion are
simply the geodesics on $\cM_k$. 

To proceed with the semiclassical analysis we need to study
the quantum mechanics of (\ref{eq:seventeen}).
Let us show how a quantised spectrum of dyons emerges in the
quantum theory. For $k=1$ we have $\cM_1=\IR^3\times S^1$ and including 
various constants we have:
\beq
S={1\over 2}\int dt\left[({4\pi v\over e})\dot \vZ^2 +{4\pi\over ve^3} \dot
\chi^2\right] -{4\pi v\over e}~.
\eeq
The wavefunctions are plane waves of the form
$e^{i\vP\cdot \vZ} e^{in_e\chi}$ where
$n_e$ is an integer. In the moduli space approximation $Q_e=-ie\p_\chi$ and
we see that we have a tower of dyons with $Q_e=n_e e$. The mass of these
states can be calculated and we get 
\bea
M&=&n_e^2 ve^3/8\pi +4\pi v/e\nonu\\
&\approx&v[Q_e^2+Q_m^2]^{1/2}~,
\label{eq:nineteen}
\eea
where we have used the fact that we are assuming $e\ll 1$ in our
approximations. By generalising the argument that led to (\ref{eq:eight}) 
it can be shown \cite{coleman} that
$M\ge v[Q_e^2+Q_m^2]^{1/2}$ for all classical solutions to the
equations of motion (static dyons can be obtained if we do not work
in the $A_0=0$ gauge).
We thus see
that in the moduli space approximation the bound is saturated. 
Of course in the purely bosonic theory we are considering here
this could get higher order quantum corrections.

For $k>1$ we can perform a similar analysis on $\cM_k$, 
looking for scattering states
and bound states of the Hamiltonian in the usual fashion. This has
been pursued in the bosonic theory in \cite{gibman,schroers}. The momentum
conjugate to the coordinate on the $S^1$ gives the total electric
charge $Q_e$ of the configuration\footnote{The individual electric charge
of each monopole is not a good quantum number due to the possibility
of $W$-boson exchange which the moduli space approximation incorporates.}
and the bound states have masses
$M=v[Q_e^2+Q_m^2]^{1/2} +\Delta E$ where $\Delta E$ is the relative
kinetic energy.

We conclude this section by considering 
a renormalisable term that we can add to the 
Lagrangian (\ref{eq:one}) that plays an important role in duality:
\beq
\delta \cL= -{\theta e^2\over 32\pi^2}F^a_{\m\n} * F^{a\m\n}~.
\eeq
As it is a total
derivative it doesn't affect the equations of motion.
It is related to instanton effects and it also affects
the electric charge of dyons. Recall that the dyon collective coordinate
arose from doing a gauge transformation about the $\Phi$ axis. The Noether
charge picks up a $\theta$ dependent contribution and one finds that
$Q_e=n_ee+e\theta/2\pi n_m$ \cite{wittendyon}. 
In the moduli space approximation
this manifests itself via $Q_e=-ie\p_\chi +e\theta/2\pi n_m$. 
At this point it is convenient to rescale the fields $\{A, \Phi\}\to
\{A,\Phi\}/e$. Our combined Lagrangian then takes the simple form
\beq
\cL=-{1\over 16\pi}{\rm Im} \tau[F^2+iF*F]-{1\over 2e^2}D\Phi^2~,
\eeq
where we have introduced the complex parameter
\beq
\tau={\theta\over 2\pi}+{4\pi i\over e^2}~.
\eeq
The BPS mass formula for dyons
(\ref{eq:nineteen}) is then given by
\beq
M=v|n_e+n_m\tau|~.
\label{eq:twthree}
\eeq
Due to the rescaling, here and in the following
$v$ contains a 
hidden factor of the coupling constant $e$.

\section{$N$=4 SUPER-YANG-MILLS}

$N$=4 super-Yang-Mills theory has the maximal amount of
supersymmetry with spins less than or equal to one. 
It has vanishing beta-function and
is thought to describe a conformally invariant theory. In addition
it is supposed to exhibit $S$-duality, which we shall define
below. 
We consider $N$=4 super-Yang-Mills with arbitrary simple gauge
group $G$. It can be obtained as the dimensional reduction on a six-torus
of $N$=1
super-Yang-Mills theory in ten dimensions (see, e.g., \cite{osborn}). 
The ten-dimensional Lorentz group reduces to $SO(3,1)\times SO(6)$ and
$SO(6)$ becomes a global symmetry of the theory.
The bosonic fields in the supermultiplet come from the ten dimensional
gauge field and
consist of a gauge field
and 6 Higgs fields $\phi^I$, transforming as a {\bf 6} of $SO(6)$,
all taking values in the adjoint representation of $G$.
There are four Weyl fermions in the adjoint transforming
as a {\bf 4} of $Spin(6)$ that come from the reduction of the Majorana-Weyl
spinor in ten dimensions.
Including a $\theta$ parameter, the bosonic part of the action is
\bea
S&=&-{1\over 16\pi}{\rm Im}\int\tau{\rm Tr}(F\wedge F+i*F \wedge F)\nonu\\
&-&{1\over 2e^2}\int \left[{\rm Tr}
D_{\mu} \phi^I D^{\mu} \phi^I+ V(\phi^I)\right]~,
\eea
where 
the potential is given by
\beq
V(\phi^I) = \sum_{1\leq I < J\leq 6} {\rm Tr} [\phi^I,\phi^J]^2~,
\eeq
and here we have taken ${\rm Tr} T^a T^b=\delta^{ab}$.

The classical vacua of the theory are given by $V(\phi^I)=0$ or
equivalently $[\phi^I,\phi^J]=0$
for all $I,J$. In this theory, there are no quantum corrections to the
moduli space of vacuua. For generic vacuua, i.e., generic
expectation
values $\langle\phi^I\rangle$, 
the gauge symmetry is broken down
to $U(1)^r$ where $r$ is the rank of the gauge group. A given $N$=4
theory is specified by $G$, $\langle\phi^I\rangle$ and $\tau$.

The six Higgs fields define a set of conserved 
electric and magnetic charges which
appear as central charges in the $N$=4 supersymmetry algebra:
\bea
Q^I_e &=& {1\over ev}\int d{\bf S} \cdot {\rm Tr}({\bf E} \phi^I)~, \nonu\\
Q^I_m &=& {1\over ev}\int d{\bf S} \cdot {\rm Tr}({\bf B} \phi^I) ~,
\eea
For BPS saturated states, i.e., states in the short 16 dimensional
representation of the
supersymmetry algebra, the mass is exactly given by the formula
\beq
M^2 = {v^2\over e^2}\bigl( (Q^I_e)^2 + (Q^I_m)^2\bigr)~.
\label{eq:tseven}
\eeq
The spin content of the short BPS multiplet is
the same as the massless multiplet and has spins $\le1$.
There are also medium
sized representations consisting of 64 states with spins $\le3/2$, 
but these only arise  when
$Q_e^I$ is not proportional to $Q_m^I$. These can only appear
when the rank of the gauge group is greater than one and
we will see that $S$-duality makes
no predictions about the existence of these states as they don't appear
in the perturbative spectrum. 
The generic representation of the $N$=4
algebra has
256 states with spins $\le 2$ and the masses can be renormalised.

It is important to emphasise that the mass 
formula for BPS states (\ref{eq:tseven})
is derived from the supersymmetry algebra and hence it is valid in
the quantum theory \cite{witol,osborn}
in contrast to the bosonic case. 
Thus the mass of BPS states is exactly given by
their electric and magnetic quantum numbers. This is an important
property of BPS states which enables us to use them to test 
$S$-duality. It will also be useful to note that half of the supersymmetry
generators are realised as zero on a BPS multiplet. This is sometimes
rephrased as saying that BPS states
preserve (or break) half of the supersymmetry.


In a generic vacuum $\langle\phi^I\rangle$ at weak coupling we deduce that
there are massive $W$-boson BPS multiplets. To determine
the dyon spectrum we need to quantise 
the BPS monopole solutions in a semiclassical context.
For simplicity we will restrict our attention in the following
to a single direction
in the moduli space of vacuua:
\bea
&&\langle\phi^2\rangle=\dots=\langle\phi^6\rangle=0~,\nonu\\
&&\phi^1\equiv\Phi~,\qquad \langle{\rm Tr}\Phi^2\rangle=v^2,
\label{eq:thirty}
\eea
which clearly satisfies $V(\phi^I)=0$. Classical BPS monopole solutions
with zero electric charge are then obtained by solving the Bogomol'nyi
equations we considered before
\beq
B_i =
D_i \Phi~.
\label{eq:tweight}
\eeq
Note that for the vacuua (\ref{eq:thirty}) only the first component
of the electric and magnetic charges are non-zero and we will write
$Q_e^1=Q_e$, $Q_m^1=Q_m$.
More general vacuua have been considered in \cite{fraserhol}
but the region we will analyse seems to lead to the
richest monopole physics.

\section{N=4 G=SU(2) AND S-DUALITY}

We now restrict our attention to gauge group $G=SU(2)$. Since
we are focusing on a single Higgs field (\ref{eq:thirty})
we will be able to directly use 
many of the results in section two. We will assume $v^2\ne 0$ so that
$SU(2)\to U(1)$. BPS states with charges $(n_m,n_e)$
satisfy the mass formula (\ref{eq:twthree}).
It is an important fact that BPS states with $(n_m,n_e)$ relatively
prime integers are absolutely stable for all values of
$\tau$. This is deduced by charge conservation and the triangle 
inequality.

We now state the $S$-duality conjecture: the $SL(2,\IZ)$
transformations
\bea
\tau&\to&{a\tau+b\over c\tau+d}~,\nonu\\
(n_m,n_e)&\to&(n_m,n_e)\left(\matrix{a&b\cr c& d\cr}\right)^{-1}~,
\label{eq:thtwo}
\eea
where $a,b,c,d\in \IZ$, $ad-bc=1$, give the same theory \cite{sen}.
The $SL(2,\IZ)$ group is generated by $T:\tau\to \tau+1$ which
is equivalent to the transformation $\theta\to\theta +2\pi$ that
can be deduced in perturbation theory (after a relabeling of states)
and $S:\tau\to -1/\tau$, which for $\theta=0$, is equivalent to strong-weak
coupling and electric-magnetic duality originally considered in 
\cite{mo,osborn}.

A simple check of S-duality is that 
the BPS mass formula (\ref{eq:twthree})
is invariant. This is a necessary condition 
because the BPS mass formula can be derived from the supersymmetry
algebra and hence it holds in the quantum theory. To see the invariance 
one should note that $v\to v'=v|c\tau +d|$ under a $SL(2,\IZ)$ transformation 
because we rescaled the higgs field by a factor of the coupling constant $e$.

We now argue that there are more sophisticated tests of
$S$-duality.
We begin by noting that the perturbative spectrum can be determined
at weak coupling and consists of
a neutral massless photon multiplet $(0,0)$ and
massive $W^\pm$-boson BPS multiplets with charge $(0,\pm 1)$.
$S$-duality maps the $W$-boson multiplets 
to BPS states $(k,l)$, with $k$ and $l$
relatively prime integers, typically at strong coupling. 
But
since these are precisely the absolutely stable BPS states they cannot decay
as we vary $\tau$ and we deduce that they must also exist at weak coupling
where we can search for them using semi-classical techniques. We will
argue that they can be translated into the existence of certain
geometric structures on the moduli space $\cM_k$.

If we assume that the entire spectrum of BPS states does not vary
as we change the coupling then we can deduce that the above BPS states are the
only BPS states in the theory. Any extra states would necessarily be
BPS states at threshold, i.e., at threshold to decay into other BPS states.
For example the mass of a potential BPS state $(2,2)$ is only
marginally stable into the decay of two $(1,1)$ states. 
If there were such 
states at threshold then we could use $S$-duality
to map them to purely electrically charged
states $(0,n)$ with $n\ne \pm 1$. 
Using our assumption that the spectrum of BPS states doesn't change
as we vary the coupling we conclude that these states should exist at 
weak coupling but this contradicts what we see in perturbation theory.
We believe that the additional assumption is weak due to the very strong
constraints that $N$=4 supersymmetry imposes on the quantum theory.
What is now known about the BPS spectrum, and will be reviewed
below, supports this assumption\footnote{See \cite{ferrari}
for an alternative way of determining the
BPS spectrum that also bears on this issue.}. 


Let us translate the prediction of the spectrum of
BPS states with relatively prime
charges $(k,l)$ into statements about the moduli space of monopoles.
The semiclassical analysis begins with the moduli space $\cM_k$ of BPS
monopole solutions. We have noted that the $4k$ coordinates on $\cM_k$
can be interpreted as collective coordinates that must be introduced
for $4k$ bosonic zero modes. In the $N$=4 context we also have fermionic
zero modes. These arise from solving the Dirac equation for the fermion
fields
in the presence of a given monopole solution. There are four Weyl or
two Dirac spinors in the adjoint of $SU(2)$ and an 
index theorem \cite{callias} tells us that
there are $4k$ fermionic $c$-number zero modes
that require the
introduction of $4k$ complex Grassmann odd fermionic ``collective
coordinates" $\psi^\alpha$. 
This means the low-energy ansatz for the fermions
will include terms of the schematic form
\beq
\lambda(x,t)\sim \psi(t)\lambda^{cl}(x,Z(t))
\eeq
where $\lambda^{cl}(x,Z)$ is a $c$-number fermion zero mode
for the monopole solution specified by the moduli $Z$.
We noted above that 
BPS states preserve half of the supersymmetry. This manifests
itself in the fact that half of the supersymmetry generators leave
the classical
BPS monopole solution invariant.
It can be shown that
the bosonic and fermionic zero modes form a multiplet of the unbroken
supersymmetries. This is essential in obtaining a supersymmetric
low-energy ansatz for the fields.
The ansatz for the low-energy fields is technically quite involved
and has been carried out in \cite{gauntlett,blum}. 
The result of substituting the ansatz into the
spacetime Lagrangian leads to the following
supersymmetric quantum mechanics
\bea
&&S={1\over 2} \int dt\big(\cG_{\alpha\beta}[\dot Z^\alpha \dot Z^\b
+i\bar\psi^\a \gamma^0D_t\psi^\b] \nonu\\
&&\quad\qquad+{1\over 6} R_{\a\b\gamma\delta}\bar\psi^\a\psi^\gamma
\bar\psi^\beta\psi^\delta\big)
-{4\pi v\over e^2 }k~,
\label{eq:thfour}
\eea
where we have traded the complex $\psi^\alpha$ for a
real two component Majorana spinor $\psi^\alpha_i$ and the covariant derivative
of these fermions is obtained using the pullback of the Christoffel 
connection: $D_t\psi^\alpha=\dot\psi^\a+\Gamma^\a_{\b\gamma}\dot Z^\beta
\psi^\gamma$.
For a general metric the supersymmetric quantum mechanics has $N$=1
supersymmetry specified by a real two component spinor 
$\eps$. In the case that the target is 
\hK\ there are an additional three supersymmetries
with parameters $\eps^{(m)}$ \cite{n4}.
Since the monopole moduli spaces are \hK\ there
are eight real supersymmetry parameters which precisely
correspond to the half of the
spacetime supersymmetry that is preserved by BPS states.

The quantisation of this model is discussed in \cite{witsusybreak}.
The states are in one to one correspondence with differential forms on
$\cM_k$. There are four real two component supercharges. Replacing
one of these with a complex one component charge $Q$ we can 
write the Hamiltonian as
$H=\{Q,Q^\dagger\}+4\pi n_mv /e^2$ where we have included the topological term.
The supersymmetry charge $Q$ is realised as the exterior derivative acting
on forms, $Q=d$, and $Q^\dagger$ as its adjoint $Q^\dagger=d^\dagger=*d*$ with
$*$ being the Hodge star acting on forms. As a consequence,
the Hamiltonian is the Laplacian acting on differential
forms
\beq
H=dd^\dagger +d^\dagger d + {4\pi v\over e^2}n_m.
\eeq

For $n_m=1$ we have $\cM_1=\IR^3\times S^1$. A basis of forms is
given by $\{1, dZ^\alpha,\dots,dZ^1\wedge dZ^2
\wedge dZ^3\wedge dZ^4\}$ which gives 16 states
corresponding to a BPS multiplet. To be more precise we need to
check that the spins of these states are the same as those of the
BPS multiplet. For $n_m=1$ all of the fermionic zero modes
can be constructed explicitly as Goldstinos by acting with the
broken supersymmetry generators. One can check the angular
momentum content and one finds that the spin content is that of a BPS
multiplet \cite{osborn}. 
The wave functions multiplying these forms are just as in the
bosonic case, $e^{i\vP\cdot\vZ}e^{in_e\chi}$, corresponding 
to dyons with $Q_e=n_ee+e\theta /2\pi$.
The Laplacian on $\IR^3\times S^1$ is trivial and by following the same
arguments as in the bosonic case, 
we deduce that the mass of these states
is given by (\ref{eq:twthree}).
Putting this together we deduce that for $n_m=1$ there is a tower of
BPS dyon states $(n_m,n_e)=(1,n_e)$ exactly as predicted by duality.
 
Now we turn to $n_m=k>1$. In this case $\cM_k=\IR^3\times(S^1
\times \tilde {\cM}^0_k)/\IZ_k$. If we first ignore the $\IZ_k$ 
identification then 
the states are 
tensor products of forms on $\IR^3\times S^1$ with forms on $\tilde \cM^0_k$,
respectively, $|s\rangle=|\omega\rangle_{n_e}\otimes |\alpha\rangle$. 
The analysis for the states 
$|\omega\rangle_{n_e}$ is similar to the $n_m=1$ case: 
there are 16 differential 
forms that are again associated with
Goldstinos and these make up the spin content of a BPS multiplet.
The wave functions give
rise to quantised electric charge
with $Q_e=n_ee+e\theta n_m/2\pi$.
The energy of the states  $|s\rangle$ can be determined
and we find
\beq
H|s\rangle=({P^2\over 2M}+M)|\omega_{n_e}\rangle\otimes |\alpha\rangle
+|\omega\rangle_{n_e}\otimes|\Delta\alpha\rangle
\eeq
with $M$ given by the BPS mass formula (\ref{eq:twthree}).
Thus to get a BPS state with charges $(n_m,n_e)$ we need
a normalisable (i.e., $L^2$) harmonic\footnote{Note that if 
$\Delta \alpha=\epsilon \alpha$ with
$\eps\ne0$ then by acting with the
supersymmetry charges one can show that it always comes
in multiplets of 16. Combining this with the 16 states $|\omega\rangle$ gives
rise to a 256 multiplet of $N$=4 supersymmetry. These states are
relevant for studying the scattering of BPS states.}
form $\alpha$ on 
$\tilde {\cM}^0_k$.
The action of $\IZ_k$ on the $S^1$ is a cylic shift which leads to 
the action $|\omega\rangle_{n_e} \to e^{2\pi i n_e/k}|\omega\rangle_{n_e}$. 
Hence for the
state $|s\rangle$ to be well defined on $\cM_k$ we need the form 
$|\alpha\rangle$ to transform 
as $|\alpha\rangle\to e^{-2\pi in_e/k}|\alpha\rangle$. 

Recalling that duality
predicts that for each relatively prime integers $(k,l)$ there is a 
unique BPS state, we conclude that $\tilde {\cM}^0_k$ must have a unique 
normalisable harmonic form which picks up a phase $e^{-2\pi il/k}$
under the $\IZ_k$ action, for every relatively prime pair of integers $(k,l)$.
The uniqueness 
implies that the form must either be self-dual or anti-self dual,
since otherwise acting with the Hodge star $*$ would generate another
harmonic form with the above properties. This conjecture was formulated
by Sen who also found the harmonic form for $k=2$.
\cite{sen}. For $k>2$ substantial evidence was provided in 
\cite{segal} (see also \cite{porrati}). 

\section{HIGHER RANK GAUGE GROUPS}
We now turn to $N$=4 theories with simple gauge groups $G$ with rank
$r>1$ with maximal symmetry breaking to $U(1)^r$. 
For simplicity of notation we will often discuss the case
$G=SU(3)\to U(1)^2$. The Lie algebra of $G$ has
a maximal abelian subalgebra $H$ 
with $r$ generators $H_i$. We can define raising and lowering
operators $E_{\pm\balpha}$ that satisfy
\bea
{}[H_i, E_\balpha] &=& \alpha_i E_\balpha,\nonu\\
{}[E_\balpha,
E_{-\balpha} ]&=&\sum_{i=1}^r \alpha^i H_i~.
\eea
(a linear combination of these generators give the $T^a$ satisfying 
${\rm Tr} T^a T^b =\delta_{ab}$
that we used before).
$\balpha$ is an $r$-component root vector.
A basis of simple roots, $\bbeta^{(a)}$ ($a=1,\cdots,r$),
may be chosen such that any
root is a linear combination of $\bbeta^{(a)}$  with  integral
coefficients all of the same sign. Positive roots are those
with positive coefficients.

We continue to work with a single Higgs field $\Phi$ by restricting
our attention to the special part of moduli space (\ref{eq:thirty}).
We may choose the Cartan subalgebra such that our vacuum is specified
by $\langle\Phi\rangle= 
v \vh\cdot \vH$ with $v^2=\langle\Tr\Phi^2\rangle$.
If $\balpha \cdot \vh=0$ for some root $\balpha$ then the unbroken
gauge group is nonabelian. Otherwise, maximal symmetry breaking
occurs, and $\langle\Phi\rangle$ picks out a unique set of simple roots
which satisfy the condition $\vh \cdot \bbeta^{(a)} > 0$ \cite{weinberg}.

Since the fields are in the adjoint representation, 
the electric quantum numbers of states
live on the $r$-dimensional root lattice
spanned by the simple roots $\bbeta^{(a)}$,
\begin{equation}
{\bf q} = \sum n^e_a \bbeta^{(a)}~,
\eeq
where the $n^e_a$ are integer.
The electric charge (for $\theta=0$) is then given by
\beq
Q_e=e\vh\cdot\vq~.
\label{eq:forty}
\eeq
At weak coupling we deduce that for 
each root $\balpha$ there is a BPS $W$-boson
with $\vq=\balpha$. 
For $SU(3)$ we have $W$-bosons with
$\vn^e =\pm(1,0),\pm (0,1)$ and $\pm(1,1)$ corresponding to the 
two simple roots $\bbeta^{(1)},\bbeta^{(2)}$ and the
non-simple positive root $\bgamma=\bbeta^{(1)}+\bbeta^{(2)}$, respectively. 
From (\ref{eq:tseven}) we deduce that the $W$-bosons
corresponding to simple roots are stable, while those
corresponding to the non-simple roots are only neutrally stable.
In $SU(3)$ we have that $M_{\bgamma}=M_{\bbeta^{(1)}}+M_{\bbeta^{(2)}}$.

Magnetic quantum numbers arise from
topologically nontrivial field configurations. For any
finite energy solution the Higgs field must approach the vacuum:
let the asymptotic value along the positive $z$-axis be
$\Phi_0=v \vh\cdot \vH$ (the value in any other direction can
only differ from this by a gauge transformation).
Asymptotically we also have
\beq
B_i = {r_i \over {4\pi r^3}} G(\Omega)~,
\eeq
where $G$ is covariantly constant, and takes the value $G_0$ along
the positive $z$-axis.
The Cartan subalgebra may
be chosen so that $G_0 = \vg\cdot \vH$. For a smooth solution this quantity
must satisfy a topological quantization condition \cite{gno,engwind}
\beq
e^{i G_0} = I~.
\eeq
The solution to this equation is
\beq
\vg= 4 \pi \sum n_a^m \bbeta^{(a)*}~,
\eeq
where the $n_a^m$ are integers and the $\bbeta^{(a)*}$ are the
the simple coroots, defined as
\beq
\bbeta^{(a)*} = { \bbeta^{(a)} \over \bbeta^{(a)2} }~.
\eeq
The magnetic quantum numbers thus live
on the coroot lattice spanned by the $\bbeta^{(a)*}$.
For maximal symmetry breaking, all of the $n_a^m$
are conserved topological charges, labeling the homotopy class of the
Higgs field configuration.
For solutions of the Bogomol'nyi equations (\ref{eq:tweight})
all of the integers in $n_a^m$ have the same sign.
The topological charge ${\bf g}$ determines the magnetic charge
by the formula
\beq
Q_m={1\over e}\vg\cdot\vh ~.
\label{eq:fortyfive}
\eeq

A general dyon state may be labeled either by the electric 
and magnetic charge $r$-vectors $\vq$, $\vg$ or by the integer
valued $r$-vectors $\vn^e$ and $\vn^m$.
{}For a BPS state the mass is given by the BPS mass formula (\ref{eq:tseven})
which,
using (\ref{eq:forty}) and (\ref{eq:fortyfive}),
can be recast
in the form
\beq
M=v|(\vh\cdot\bbeta^{(a)})n^e_a+\tau(\vh\cdot\bbeta^{(a)*})n^m_a|~,
\label{eq:fosix}
\eeq
where we have reinstated $\theta$.

We now have enough definitions to define the action of $S$-duality.
It is the natural generalisation of the $SU(2)$ case (\ref{eq:thtwo}): 
the $SL(2,\IZ)$ duality on a general dyon state is given by
\bea
\tau&\to&{a\tau+b\over c\tau +d}~,\nonu\\
(\vn^m,\vn^e)&\to&(\vn^m,\vn^e)\left(\matrix{a&b\cr c& d\cr}\right)^{-1}~,
\eea
and when we act
with the $S$-generator $S:\tau\to-1/\tau$ 
we must replace the group $G$ with its dual group $G^*$
\cite{gno}. For simply laced groups i.e., all the roots have the same 
length (the $ADE$ groups), the $N$=4 supersymmetric Lagrangian with
gauge group $G$ is the
same as that of $G^*$ since all fields are in
the adjoint representation. For non-simply-laced groups
this is not true since for example $SO(2N+1)^*=Sp(N)$. In this
case one does not expect the theory to be invariant under the
full $SL(2,\IZ)$ duality group, but rather a $\Gamma_0(2)$
subgroup \cite{girard} (see also \cite{dfhk}). We restrict our considerations to simply-laced
gauge groups in the following. 

Just as in the $SU(2)$ case $S$-duality maps the perturbative $W$-boson states 
into an infinite number of dyon BPS states. For the $SU(3)$ case we
generate the following $SL(2,\IZ)$ orbits:
\bea
\bigl(\vn^m,\vn^e\bigr)=&&\bigl(k(1,0),l(1,0)\bigr), \nonu\\
&&\bigl(k(0,1),l(0,1)\bigr),\nonu\\
&&\bigl(k(1,1),l(1,1)\bigr)~,
\label{eq:foeight}
\eea
for relatively prime integers $k$ and $l$. Like the $SU(2)$ case
we have again typically been mapped to strong coupling. For the first two
classes of states we note from the BPS mass formula (\ref{eq:fosix})
that they are
absolutely stable and hence we conclude that they also exist at weak
coupling. The whole orbit of states coming from the $(1,1)$ $W$-boson
are only marginally stable. Consequently we have to again employ the
additional assumption that in the $N$=4 theory the spectrum of marginal
states does not change as we vary the coupling. In this case 
we should see these states at weak coupling also.

It is perhaps worth noting here that by starting with 
the perturbative spectrum of $W$-bosons
$S$-duality only makes
predictions about the short BPS representations of the 
$N$=4 supersymmetry algebra.
This is because the purely
electrically charged $W$-bosons have parallel
electric and magnetic charge vectors $Q^I_e$ and $Q^I_m$.
If any medium sized representations of the $N$=4 algebra existed
they would necessarily have non-parallel charge vectors and lie
on separate $SL(2,\IZ)$ orbits. It would be interesting to know if
they existed.

To test the $S$-duality predictions (\ref{eq:foeight})
we begin by reviewing some aspect of BPS monopole
solutions.
Using an index theorem Weinberg has argued that the moduli
space of monopoles of charge $\vn^m$ has dimension
\beq
d=4\sum_a n_a^m~.
\label{eq:fonine}
\eeq
A number of explicit monopole solutions can be constructed by embedding $SU(2)$
monopoles as follows \cite{bais}. Let $\phi^s$, $A_i^s$ be an $SU(2)$ monopole
solution with charge $k$ and Higgs expectation value $\lambda$. If we
let $\balpha$ be any root satisfying $\balpha\cdot\vh>0$ then we can
define an $SU(2)$ subgroup with generators
\bea
t^1 &=& (2 \balpha^2 )^{-1/2} ( E_{\balpha} +
E_{-\balpha}) \nonu\\
t^2 &=& -i(2 \balpha^2 )^{-1/2} ( E_{\balpha} -
E_{-\balpha}) \nonu\\
t^3 &=& (\balpha^2 )^{-1} \balpha \cdot \vH ~.
\eea
A monopole with magnetic charge
\beq
\vg=4\pi k\balpha^*
\label{eq:fione}
\eeq
is then given by
\bea
\Phi &=& \sum_s \phi^s t^s + v( \vh - { {\vh\cdot \balpha }\over
{\balpha^2}} \balpha ) \cdot \vH \nonu\\
A_i &=& \sum_s A_i^s t^s \nonu\\
\lambda&=&v\vh\cdot\balpha~.
\eea
Since the moduli space of $SU(2)$ monopoles with charge $k$ has dimension $4k$
these solutions provide a $4k$ dimensional submanifold of monopole solutions
with charge (\ref{eq:fione}).
Note that by embedding an $SU(2)$ monopole with charge one we obtain
spherically
symmetric monopole solutions.

Weinberg has shown that
there is a distinguished set of $r$
``fundamental monopoles" with $\vg=4\pi\bbeta^{(a)*}$ i.e., they have magnetic
charge vectors $\vn^m$ consisting of a one in the $a$th position and
zeroes elsewhere. The reason for calling them fundamental is
twofold. First, they have no ``internal" degrees of freedom:
all of these solutions can be constructed by
embedding an $SU(2)$ monopole of unit charge using
the corresponding simple
root and consequently
they have only four zero modes: three translation
zero modes and a $U(1)$ phase zero mode corresponding to dyonic excitations
of the same $U(1)$ as where the magnetic charge lies\footnote{One can
check that the embedded $SU(2)$ solutions are invariant under
gauge transformations of the other $U(1)$'s.}.
Secondly, the index theorem (\ref{eq:fonine}) is consistent with
thinking of a general monopole with charge $\vn^m$
as a multimonopole configuration consisting
of $n_a^m$ fundamental monopoles of type $a$. 

Note that for magnetic monopoles with charge vector $\vg=4\pi k\bbeta^{(a)*}$
i.e., consisting of $k$ fundamental monopoles of the same type, the dimension
of moduli space is $4k$. Thus we deduce that these solutions can all be
obtained by embedding $SU(2)$ monopoles of charge $k$, using the embedding
based on the same simple root.

Let us now return to the BPS states predicted by $S$-duality.
We need to study the semiclassical quantisation for a given
magnetic charge $\vn^m$. Just as in the $SU(2)$ case the
bosonic zero and fermionic zero modes are paired by the 
unbroken supersymmetry
and a low-energy ansatz again leads to the $N$=4 supersymmetric
quantum mechanics (\ref{eq:thfour})
on the moduli space of solutions $\cM_{\vn^m}$.
First consider monopoles with $\vn^m=(k,0)$ or $\vn^m=(0,k)$
i.e., $k$ fundamental monopoles of the same type. The moduli space
of these monopoles is the $SU(2)$ moduli space $\cM_k$. 
The dyonic states with charges $( k(1,0), l(1,0))$
and $(k(0,1),l(0,1))$ predicted by duality are equivalent to
the harmonic forms on $\cM_k$ required by $S$-duality in
the $SU(2)$ theory. The results of
\cite{sen,segal} thus constitute tests of duality for higher rank
gauge groups. 

The new predictions for $SU(3)$ monopoles arise in the sectors
with both magnetic quantum numbers non-zero. In particular, the
$(k(1,1), l(1,1))$ dyon states should arise as bound states of $k$
$(1,0)$ and $k$ $(0,1)$ monopoles. Note from the BPS mass
formula that these states are only neutrally stable and consequently
they should emerge as bound states at threshold. At present only
for $k=1$ have these states been shown to exist. Let us make
some comments on this case.

It was shown in \cite{gl,lwy,connell} that the moduli space 
for $\vn^m=(1,1)$ is given
by 
\beq
\cM_{(1,1)}=\IR^3\times {\IR\times \cM_{\rm TN}\over \IZ}
\label{eq:fithree}
\eeq
where
$\cM_{\rm TN}$ is four-dimensional Taub-NUT space. The $\IR^3$ factor
corresponds to the centre of mass of the $(1,0)$ and $(0,1)$ 
monopole configuration. Taub-NUT space
is a \hK\ manifold as required for the quantum
mechanics (\ref{eq:thfour}) to have
$N$=4 supersymmetry. Taub-NUT space has
$U(2)$ isometry, of which a $SU(2)_L$ subgroup corresponds to the action of
rotating the multimonopole configuration in space. That this is $SU(2)$
and not $SO(3)$ can be demonstrated by studying the zero modes about
the spherically symmetric $(1,1)$ solution that can be obtained via
the $SU(2)$ embedding using the root $\bgamma$ \cite{gl}. 
Note that the fixed point set of the $SU(2)_L$ action is a single point
in Taub-NUT space (the ``nut") and this corresponds to the spherically
symmetric solutions. The extra $U(1)_R$ isometry in $U(2)$ combined with
the factor $\IR$ in (\ref{eq:fithree})
and the identification under the integers $\IZ$
lead to dyon states with electric charge $\vn^e$.
One might have expected an $S^1$ factor rather than $\IR$ for the total
electric charge in the $(1,1)$ direction 
(i.e., parallel to the magnetic charge), but this is not quite correct
due to the fact that in general the masses of the two
fundamental monopoles $(1,0)$ and $(0,1)$ are not equal.

The basis of
16 forms on $\IR^3\times \IR$ leads to a BPS supermultiplet of 16 states in
the $N$=4 supersymmetric quantum mechanics. 
In order to get the 
dyon BPS states predicted by duality (\ref{eq:foeight}) with magnetic charge 
$(1,1)$
there must exist a unique
normalizable harmonic (anti)-self-dual two-form
on Taub-NUT space that is 
invariant under the $U(1)_R$ isometry 
to ensure that the
electric charge is $l(1,1)$. Such a harmonic form exists \cite{gl,lwy}.

\section{SOME OPEN PROBLEMS}

As we have discussed a number of highly non-trivial checks of
exact $S$-duality can be carried out by studying the 
geometry of monopole moduli spaces.
Let us conclude by discussing some open  
issues.

In the $N$=4 theory with maximal symmetry breaking the BPS
states predicted by duality correspond to normalisable harmonic forms on BPS
monopole moduli spaces.
For gauge group $SU(2)$ the work of
\cite{segal} provided substantial evidence for the appropriate harmonic forms.
It would be desirable to have a similar
analysis for higher rank gauge groups. For $SU(3)\to U(1)^2$ we have
seen that the duality predictions 
for monopoles with $k$ fundamental monopoles of the same type,
i.e., $\vn^m=k(1,0)$ or  $k(0,1)$ reduce to those of the $SU(2)$ case.
The new $SU(3)$ predictions with $\vn^m=(1,1)$ correspond to a harmonic
form on Taub-NUT space.
It remains to be shown that the BPS states with $\vn^m=k(1,1)$
exist for $k\ne 1$ and that the BPS states in (\ref{eq:foeight}) 
are the only ones in the spectrum. 
The $SU(2)$ and $SU(3)$
results can be embedded in higher rank gauge groups. This can be illustrated
by considering $G=SU(4)\to U(1)^3$. In this case $S$-duality 
predicts BPS states with $\vn^m=k(1,0,0)$, $k(0,1,0)$, $k(0,0,1)$, which
are equivalent to the $SU(2)$ predictions, $k(1,1,0)$, $k(0,1,1)$ which
are equivalent to the $SU(3)$ predictions, and $k(1,1,1)$ which are
the new $SU(4)$ predictions. Apart from the cases we have discussed
there is only one more class of moduli spaces that are explicitly
known: 
when there are no more than a single fundamental monopole of each type
(e.g, $(1,1,1)$ for the $SU(4)$ example)
\cite{lwytwo,murray,chalmers}. 
It is a natural generalisation of Taub-NUT
space and the harmonic form predicted by duality has been shown to
exist \cite{gibbons}.
The major obstacle in verifying more of the $S$-duality
predictions is our lack of knowledge about the monopole moduli spaces.

New issues arise in the $N$=4 case when a non-abelian gauge group
remains unbroken. In this case the existence of massless $W$-bosons
might seem to require dual massless monopoles which cannot be 
studied as conventional semi-classical solitons. There
are also massive monopoles that can be studied.
If they carry non-abelian magnetic charge
there are subtleties to do with the moduli space approximation
due to the non-normalisablility of zero modes corresponding
to global gauge rotations (see e.g., \cite{doreyetal,lwythree}). 
The moduli spaces of monopoles that have net abelian magnetic charge 
can in some cases be determined as limits of
moduli spaces in which the symmetry is maximally broken. Curiously,
it is claimed that the harmonic forms found by \cite{gibbons}
become non-normalisable
in this limit \cite{lwythree}. 

In this paper we have only discussed $N$=4 theories. Special
theories with $N$=2 supersymmetry and vanishing $\beta$-function
are also candidates for exhibiting exact duality. For
gauge group $SU(2)$ with $N_f$=4 hypermultiplets in the 
fundamental representation, it is conjectured that the duality group
is the semi-direct product of 
$SL(2,\IZ)$ with the global flavour symmetry group $Spin(8)$ \cite{swt}.
$SL(2,\IZ)$ mod 2 is
isomorphic to $S_3$ the permutation group of three objects, which is
also the group of outer automorphisms of Spin(8)
which acts on the ${\bf v},{\bf s},{\bf c},$ conjugacy classes. This duality
predicts  an orbit of vector multiplets at threshold with charges
$(n_m,n_e)=2(k,l)$ and an orbit of hypermultiplets with charges $\pm(k,l)$.
In this theory there are half as many fermionic zero modes coming
from the vector multiplet as in the $N$=4 theory and the net
result is that one should study an $N$=2 supersymmetric quantum
mechanics on the moduli space of $SU(2)$ monopoles $\cM_k$. As a consequence,
the states are spinors on $\cM_k$ not forms.
In addition there are fermionic zero modes coming from the hypermultiplets
that give rise to a natural $O(k)$ bundle on $\cM_k$ \cite{mantonschroers}. 
The BPS states
predicted by duality 
correspond to certain harmonic spinors coupled to this bundle. For 
monopole charge $k=2$ these were found using index theory 
\cite{gauntharv,sethietal}. 
Perhaps an analysis
similar to \cite{segal} is possible for higher monopole charge.
See also \cite{ferrari} for a different
approach. 

For higher rank theories with $N$=2 supersymmetry and vanishing
$\beta$-function less is known about exact duality. 
A straightforward attempt to find
the duality group acting on the lattice of electric and magnetic charges
was attempted in \cite{cederwall} for $G=SU(3)$ (see also \cite{yank})
but the results were inconclusive.

All of these tests we have been discussing
concern the spectrum of BPS states. If the exact duality 
conjectures are true then they should also apply to non-BPS states, for
example the scattering of BPS states. Since such process are not protected by 
supersymmetry it remains a challenging problem to find evidence for
$S$-duality in this sector.

More generally we would like to know the underlying reasons 
for duality in field theory.
String theory duality would seem to provide one answer since we can
embed these gauge theories in various string theory settings. Of course
this still leaves the more involved issue of elucidating 
the deeper principles that underly
string theory duality.

%

\end{document}